\documentstyle[epsfig,rotate,12pt]{article}
\begin{document}
\newcommand{\be}{\begin{equation}}
\newcommand{\ee}{\end{equation}}
\newcommand{\diff}[2]{\frac{\partial #1}{\partial #2}}
\newcommand{\con}[2]{\multicolumn{#1}{c}{} &
            \multicolumn{#2}{c}{\vrule width\arrayrulewidth depth0mm
            \hrulefill \vrule width\arrayrulewidth depth0mm}&}
\newcommand{\nl}{\\[-1.6ex]}
\newcommand{\contract}[3]{{
            \renewcommand{\arraystretch}{0.7}
            \hskip-0.4em\begin{array}[t]{*{#1}{c@{\hskip0em}}c}
            #2 \\[-0.7ex] #3 \\ \end{array}\hskip-0.4em}}
\newcommand{\trivcontract}[1]{\contract{4}{#1}{\con{1}{3}}}
\newcommand{\doublecontract}[1]{\contract{8}{#1}{\con{1}{3}\con{1}{3}}}
\newcommand{\crosscontract}[1]{\contract{8}{#1}{\con{1}{5}\nl\con{3}{5}}}
\newcommand{\nestcontract}[1]{\contract{8}{#1}{\con{3}{3}\nl\con{1}{7}}}
\title{RPA-Approach to the Excitations of the Nucleon, Part II: Phenomenology}
\author{ S. Hardt, J. Geiss, H. Lenske and U. Mosel \\
Institut f\"ur Theoretische Physik \\ Universit\"at Giessen \\
Heinrich-Buff-Ring 16 \\ D-35392 Giessen, Germany}
\maketitle
\begin{abstract}
The tensor-RPA approach developed previously in part I is applied
to the Nambu-Jona-Lasinio (NJL) model. As a first step we investigate
the structure of Dirac-Hartree-Fock solutions for a rotationally
and isospin invariant ground-state density. Whereas vacuum properties
can be reproduced, no solitonic configuration for a system with
unit baryon number is found. We then solve the tensor-RPA equation
employing simple models of the nucleon ground state.
In general the $ph$ interaction effects a decrease of the excited
states to lower energies. Due to an enhanced level density at low
energies the obtained spectra cannot be matched 
with the experimental data when a standard MIT-bag configuration
is used. However, when the size of the nucleon quark core is reduced
to $ \approx 0.3 fm $ a fair description of the baryon spectrum
in the positive-parity channel
is achieved. For this purpose the residual interaction
turns out to be crucial and leads to a significant improvement
compared with the mean-field spectra. 
\end{abstract}
\section{Introduction}
In a previous paper \cite{HGLM96} we have developed a method to compute
the excited states of the nucleon in the framework of a
Dirac-RPA description applied to an effective quark model.
The method is based on the tensor coupling of particle-hole ($ph$)
configurations on the nucleon ground state and allows to construct
excited states with good angular momentum ($J$) and isospin ($T$)
quantum numbers. \\
In the present work we apply the tensor-RPA formalism to
a simple quark model of Nambu-Jona-Lasinio (NJL) type.
Our investigations are analogous to extensions of the NJL soliton model
beyond the mean-field level \cite{AARW95}, with two major differences. \\
First of all we do not use an Euclidean path integral formalism to
derive many-body approximation schemes, but rather rely
on the Dirac-RPA method developed in part I of this paper.
This approach is formally more similar to the conventional
formulation of nonrelativistic many-body theory. 
As a consequence, we are spared the problem of continuing back
the vector potentials to Minkowski space which has turned out
to be a quite nontrivial task \cite{SDAG93,ARW95}. \\
Secondly we treat the constraints set by the symmetries of
the interaction in a stricter way than it is done in the
NJL soliton model. In the latter states with good $J/T$
quantum numbers are projected out from hedgehog configurations
by perturbative cranking \cite{ARW95,CGGP95}. In contrast to
that we construct a ground-state density matrix which is
invariant with respect to rotations in coordinate and isospin space.
In this way we obtain a spectrum of single-particle states carrying
good $J/T$ quantum numbers which can be used to build collective
baryonic states by means of the tensor-RPA techniques. \\
The strict treatment of symmetries has important implications
for the structure of the mean field. Whereas in a simple
version of the NJL soliton model the scalar field is accompanied
by a pseudoscalar mode \cite{ARW95}, the ground state configuration 
of our model does not allow for such a mode. 
Without the pseudoscalar field the ground state as a self-consistent
solution of a regularized Hartree-Fock equation turns out to 
be an unbound system of quarks. \\  
An issue to be addressed in the context of a model for
the excited states of the nucleon is the question of confinement.
Due to the lack of a confinement mechanism 
unphysical single-quark thresholds appear in the NJL soliton model
\cite{AARW95} as well as in NJL-based approaches relying on
the Bethe-Salpeter equation in the quark-quark \cite{BR95}
or quark-antiquark channel \cite{KLVW90}. In the NJL soliton model
all excited single-quark states are unbound.
While this might be no serious drawback as long as we consider
the nucleon ground state, the description of the excitation
spectrum becomes questionable in such a description. \\
There exist models for baryons \cite{IK78} as well as for mesons
\cite{UK95} which incorporate the confinement property of
QCD, but the origin of these models lies in a nonrelativistic
potential picture for the two-body interaction. In that
framework it seems impossible to describe the structure of the
QCD vacuum and the spontaneous breakdown of chiral symmetry. \\
To our knowledge so far nobody has formulated a consistent theory
based on quark degrees of freedom exclusively
which confines a system of three quarks on mean-field level 
and models the quark sector
of the QCD vacuum at the same time. However, at the price of introducing
boson fields that couple to the quarks this goal can be reached,
like in the chromodielectric model \cite{FPW88}.
A few attempts towards a unified description of vacuum properties
and quark confinement have been made \cite{FT91,BK92,BB95}, 
but still these models suffer from a number of drawbacks. 
To give an example, an interaction of the
form $ \delta^{(4)} (p)$ as in Ref.\ \cite{FT91}
incorporates confinement in the
sense of a pole-free quark propagator, but cannot be
consistently used for finite systems as the nucleon.
Recently a confining model has been proposed which contains a remnant
of the functional integration over gluon degrees of freedom
by a statistical treatment of the $4$-quark vertex \cite{LR95}.
However, the applicability of this model to baryonic systems
is yet to be proven. \\
For applying the tensor-RPA scheme it is necessary
to model the nucleon by a mean-field configuration from which
at least some gross properties of the excitation spectra, e.\ g.\
the absence of free quarks, can be derived.
With no suitable mechanism for dynamical quark confinement at hand
we have to restrict ourselves to a simplified description of the
ground state. 
For this purpose we use a phenomenolocigal basis
of single-particle wavefunctions taken from a bag-model
together with a residual interaction of NJL type.
The bag model clearly does not represent a self-consistent
configuration in such a framework but provides us
with a simple way to incorporate confinement in our
description. \\
The present work is divided into two major sections. \\
In section 2 we will first discuss self-consistent mean-field configurations
of the vacuum. We further investigate the
structure of the solutions when we add three valence quarks to
the system. These investigations within the NJL model
are based on the most general
form of the self-energy compatible with the symmetries 
of the ground state. Subsequently we briefly discuss a class
of phenomenological models for the nucleon. \\
In section 3 we focus our attention on the excitation spectrum
as obtained with the tensor-RPA method. We compare different
models for the ground state as well as different forms of the residual
interaction. We discuss the effects of a pion-exchange interaction
which was recently claimed to be of special importance
for the baryon spectrum \cite{GR95}. Using a local color-current
interaction and reducing the size of the quark core to
$ \approx 0.3 fm $ the excitation spectrum in the positive-parity
channel can be reproduced fairly well.
\section{The Nucleon Ground State}
The major part of this section is devoted to an investigation
of self-consistent solutions of the NJL model on mean-field level.
Solitonic field configurations of the NJL model in the sector 
with unit baryon number ($B$$=$$1$) 
have been studied extensively in the past \cite{ARW95,CGGP95},
but the prescription used in these studies to project on 
physical states with good angular momentum (J) and isospin (T)
is rather crude. The tensor-RPA formalism developed in part I
constitutes a powerful tool to investigate
the excitation spectrum of the nucleon in any $J/T$ channel.
It is therefore worthwhile to explore how far the NJL model is suited 
to provide a ground-state configuration for the tensor RPA. \\
We will find that in a model space of MIT-bag states with
an appropriately large bag radius vacuum properties as obtained
from the gap equation can be reproduced with high accuracy, but
the $B$$=$$1$ system does not emerge 
as a bound solitonic object. \\
With focus on a description of the excitation
spectrum we proceed to discuss a class of simple models
for the nucleon ground state, thus giving up self-consistency
but introducing quark confinement on a phenomenological level.
       
\subsection{Self-Consistent Mean-Field Configurations}
As a model for low-energy QCD we consider the well known
NJL Lagrangian restricted to flavor $ SU(2) $ (see, e.\ g., \cite{VW91})
\be
{\cal L} = \bar{\Psi} (i \partial \! \! \! / - \hat{m}_0 ) \Psi
- G j_{a \mu} j^{\mu}_a ,
\label{colcu}
\ee 
where a color-current interaction with
\be
j^{\mu}_a = \bar{\Psi} \frac{ \lambda_a }{2} \gamma^{\mu} \Psi 
\ee
is used.
We have introduced a $SU(2)$ current-quark mass matrix $ \hat{m}_0 $
and denoted the generators of the color group by $ \frac{ \lambda_a }{2} $.
The model defined in Eq.\ (\ref{colcu}) is chirally invariant 
in case of a vanishing mass matrix $ \hat{m}_0 $. \\
To compute exchange terms we take the Fierz transform of the
color-current interaction in Eq.\ (\ref{colcu}) and obtain
\begin{eqnarray}
\lefteqn{{\cal F} \left[ G \left( \bar{\Psi} \gamma_{\mu} 
\frac{ \lambda_a}{2}
\Psi \right)^2 \right]} \nonumber \\
& & = \frac{G_s}{9}  \left[
- \left( \bar{\Psi} \Psi \right)^2 + \left( \bar{\Psi} \gamma_5
\Psi \right)^2 
- \left( \bar{\Psi} \tau \Psi \right)^2 
+ \left(  \bar{\Psi} \gamma_5 \tau \Psi \right)^2 \right] \nonumber \\
& & \; + \frac{G_v}{9} \left[ \left( \bar{\Psi} \gamma_{\mu} \Psi \right)^2
- \left( \bar{\Psi} \gamma_5 \gamma_{\mu} \Psi \right)^2
+ \left( \bar{\Psi} \gamma_{\mu} \tau \Psi \right)^2
- \left( \bar{ \Psi} \gamma_5 \gamma_{\mu} \tau \Psi \right)^2 
\right] +c.o. \; ,
\label{fierz} 
\end{eqnarray}
with
\be
G_v = \frac{G_s}{2} = G . 
\label{couplings}
\ee  
The color octet terms, indicated by $c.o.$, are not displayed, because they
do not play any role for matrix elements taken between color-singlet
states which are the only terms of interest.
The evaluation of exchange terms of the NJL-interaction is
done by computing the direct contribution of the Fierz transform. 
The interaction of Eq.\ (\ref{fierz}) is only used in the direct
channel and clearly exhibits the Lorentz and isospin structure
of the exchange matrix elements. \\ 
Chiral invariance is preserved if
we treat $G_s$ and $G_v$ as independent parameters, so in
the following we relax the condition in Eq.\ (\ref{couplings})
and consider a more general class of models with $ G_s \neq 2 G_v$. 
There are some attempts to fix the ratio $ \frac{G_v}{G_s} $
from phenomenology \cite{CGP95,YZ94}, but so far the results 
are contradictory. \\
\subsubsection{The $ B = 0 $ vacuum sector}
\begin{figure}
\vspace{-2cm}
\centerline{\rotate[r]{\epsfig{file=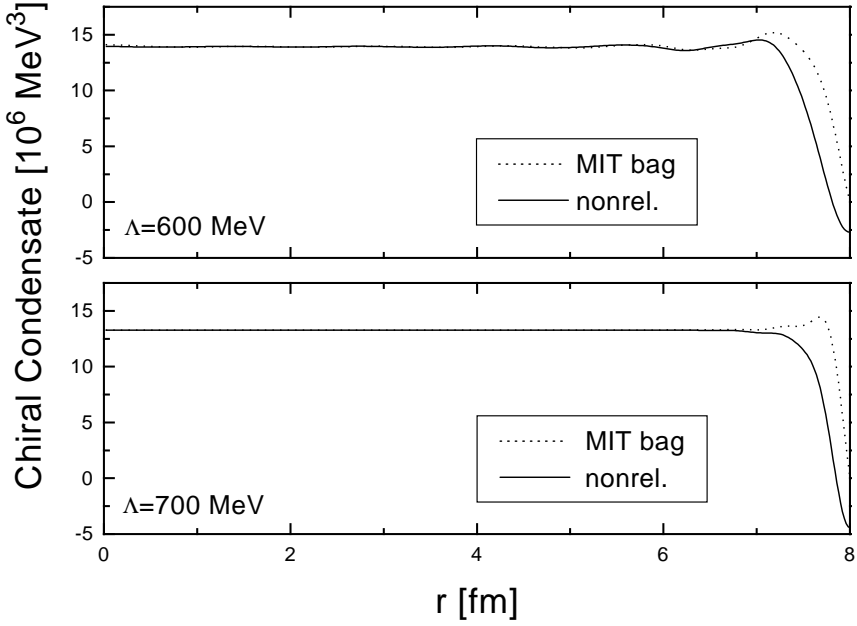,height=13cm}}}
\caption{ $ - \langle \bar{\Psi} \Psi \rangle $ for the Woods-Saxon
(above) and the proper-time (below) cut-off scheme. In both
cases $ \Lambda $ and $G_s$ were fixed to produce a
constituent-quark mass of $400 MeV$. Different boundary conditions
are compared.}
\label{conds}
\end{figure}

\begin{figure}
\vspace{-2cm}
\centerline{\rotate[r]{\epsfig{file=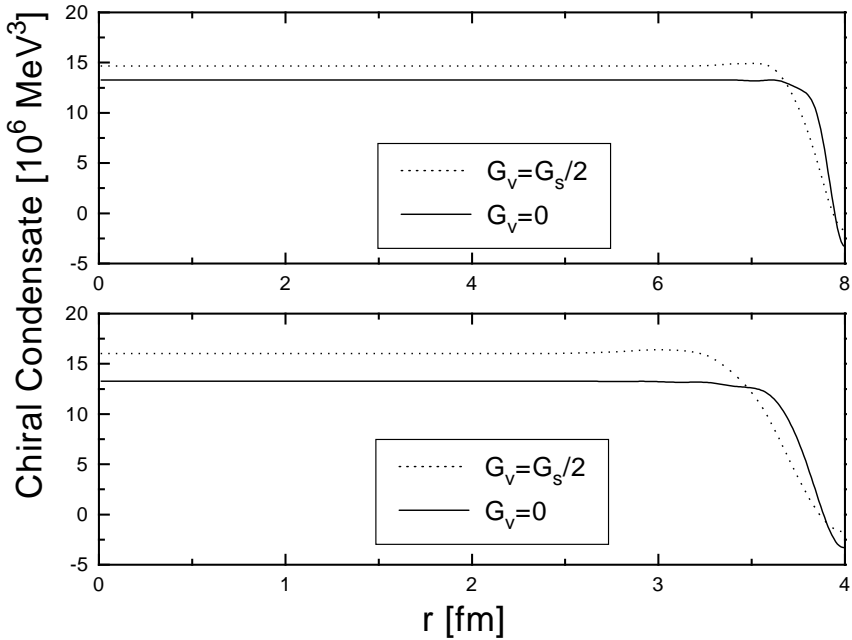,height=13cm}}}
\caption{$ - \langle \bar{\Psi} \Psi \rangle $ in a cavity with
$R_c=8fm$ (above) and $R_c=4fm$ (below)}.
\label{condv}
\end{figure}

First we investigate the vacuum-structure obtained
from the NJL model on mean-field level.  
Starting from the color-current
interaction Eq.\ (\ref{colcu}) the Hartree contribution to
the self-energy vanishes for a color-singlet vacuum. 
The Fock term is computed
by evaluating the direct contribution from the Fierz transform.
Assuming a Poincar\'e-invariant vacuum, the only nonvanishing
contribution to the self-energy is a scalar term 
\be
\Sigma_s = - \frac{2}{9} G_s \langle \bar{\Psi} \Psi \rangle.
\label{gap}
\ee
The scalar self-energy, which is up to a factor 
the chiral condensate of the QCD vacuum,
can be computed by solving a gap equation,
as first shown by Y.\ Nambu and G.\ Jona-Lasinio in 1961 
\cite{NJL61}. For flavor $SU(2)$ the chiral condensate is
given by
\be
\langle \bar{\Psi} \Psi \rangle =
- \frac{6}{\pi^2} \int_m^{\infty} d E \sqrt{ E^2 -m^2 } m R^2 [E],
\ee 
with
\be
m= m_0 + \Sigma_s, \; \; m_u=m_d=m_0.
\ee
The cut-off function $ R$ is chosen as in part I.
With the expression for the chiral condensate Eq.\
(\ref{gap}) represents a nonlinear equation for the scalar self energy,
which can be solved iteratively. \\
We take up the problem of vacuum structure in the NJL model
and tackle it from a different angle. The Dirac Hartree-Fock problem
for the nucleon will be solved in a basis of MIT-bag states, therefore
it is advisable to first perform the corresponding calculation
for the vacuum in the same basis. 
By comparing the results with the solution of the gap equation we
get an estimate on the importance of finite-size effects
in a basis corresponding to a certain bag radius.
Furthermore it allows us to make a statement about the stability
of a translationally invariant vacuum state, which enters
the gap equation Eq.\ (\ref{gap}) only as an assumption. \\
Let us consider the mean-field potential of the
vacuum in a similar way as the corresponding
expression for the nucleon we have given in part I
of this paper. From the Fierz transform Eq.\ (\ref{fierz})
we obtain the Lorentz-scalar contribution to the mean field as
\be
V^s_{MF} =  - \frac{2}{9} G_s \sum_{pqi} \left( \int d^3 r 
\bar{\Psi}_p \Psi_q \bar{\Psi}_i \Psi_i \right) 
a^{\dagger}_p a_q R^2 [\epsilon_i ] \theta ( - \epsilon_i ) .
\label{scalar} 
\ee
The time-like component of the Lorentz-vector self-energy reads
\be
V^v_{MF} =  \frac{2}{9} G_v \sum_{pqi} \left( \int d^3 r 
\Psi^{\dagger}_p \Psi_q \Psi^{\dagger}_i \Psi_i \right) 
a^{\dagger}_p a_q R^2 [\epsilon_i ] \theta ( - \epsilon_i ) .
\label{vector} 
\ee
The sums in these expressions are over the self-consistent
one-body states.
The single-particle energies $ \epsilon_i $ 
are the eigenvalues of the mean-field
Hamiltonian including the kinetic-energy contribution.
The form of the regularization function $ R $ is yet to be
specified. \\
Since the vacuum is a ($J^p$$=$$0^{+}$,$T$$=$$0$)
state, there are no contributions to the
mean field from isovector, pseudoscalar, $3$-vector or axial-vector terms.
A nonvanishing time-like component of a four-vector breaks
Lorentz invariance. Hence the potential in Eq.\
(\ref{vector}) should have no influence on the dynamics of the system.
However, besides the scalar condensate 
the vector potential is the only nonvanishing contribution
to the mean field in the corresponding problem for
the nucleon. Therefore, by inclusion of the vector potential
in the vacuum calculation we study the {\em fluctuations} of the
baryon density while keeping the effective baryon number zero
by means of a Lagrange multiplier. \\
For this purpose we consider the constrained Hartree-Fock problem
\be
V^v_{MF} \longrightarrow V^v_{MF} - \lambda \frac{2}{9} G_v B,
\label{consthf}
\ee
where the baryon number operator is given by
\be
B =
\frac{1}{3} \sum_{pq} \left( \int d^3 r \Psi^{\dagger}_p \Psi_q \right)
a_p^{\dagger} a_q .
\ee
The Lagrange multiplier $ \lambda $ is determined such that the
effective baryon number
\be
\frac{1}{3}\left( \int d^3 r \left( \sum_i \Psi^{\dagger}_i \Psi_i
R^2 [ \epsilon_i] \theta (-\epsilon_i) \right) - \frac{1}{3} \lambda \right)
\ee
vanishes. \\
The expressions for the mean-field potential, Eqs.\ (\ref{scalar})
and (\ref{vector}), have been given in the self-consistent basis
where only the diagonal elements of the regularized
mean-field Hamiltonian enter as an argument of the cut-off function
$R$. In a practical calculation we start with a field configuration
in the vicinity of the self-consistent solution, but
according to the prescription for the Hartree-Fock iteration scheme
given in part I of this paper we assume that the off-diagonal
elements of the mean-field Hamiltonian
can be neglected in $R$, so that the operators we have to
evaluate are of the same structure as above. \\
Owing to the symmetry of the vacuum the single-particle
wavefunctions can be written as 
\be
\Psi_p = \left( \begin{array}{c} g_p \chi^{\mu_p}_{\kappa_p} \\
-i f_p \chi^{\mu_p}_{-\kappa_p} \end{array} \right) , 
\ee 
where $ \chi^{\mu}_{\kappa} $ are the spin-angle functions (see,
e.\ g., \cite{Sak67}) with angular-momentum projection
$ \mu $ . $ \kappa $ denotes the eigenvalue of
$ K = \beta ( \mbox{\boldmath$\Sigma$\unboldmath} \cdot {\bf l}
+ 1 ) $. 
The radial functions $g,f$ are to be determined
self-consistently. \\
To solve for the self-consistent states, we write the mean-field
Hamiltonian in a basis of eigenmodes of a spherical cavity with
radius $R_c$ 
\be
\Psi = N \left( \begin{array}{c} j_l ( pr) \chi^{\mu}_{\kappa} \\
i \frac{p}{E+m} sgn(\kappa) j_{\bar{l}} (pr) \chi^{\mu}_{- \kappa}
\end{array} \right) ,
\ee
where $m$ is the quark mass and $ E= \sqrt{{\bf p}^2 +m^2}$.
The eigenvalues $p$ are fixed by the boundary condition. In general
the structure of the field configuration
which can be represented with a set of basis states depends
on the boundary condition imposed on these states \cite{ARSW94}. We
consider the two different forms
\begin{itemize}
\item[1.] MIT-bag boundary condition
\be
{\bf n} \cdot {\bf J} |_{R_c} = 0,
\label{mitbound}
\ee
where ${\bf n} $ is a unit vector normal to the bag surface and
$ {\bf J} = \bar{\Psi} \mbox{\boldmath$\gamma$\unboldmath } \Psi $.
\item[2.] nonrelativistic boundary condition 
\be
j_l (pR_c) = 0 .
\ee
\end{itemize}
Both conditions lead to a set of orthogonal basis states. \\
For the mass parameter $m$ we choose a value consistent
with the solution of the gap equation Eq.\ (\ref{gap}), where $ m_0 = 5 MeV$.
A set of eigenmodes of the spherical cavity with a mass parameter
equal to the constituent-quark mass from the
gap equation also serves as an initialization for our
iteration scheme. \\
We consider two different forms for the cut-off function $R$:
\begin{itemize}
\item[1.] Woods-Saxon cut-off
\be
R^2[x] = \left( 1 + e^{- \frac{x+ \Lambda}{\sigma}} \right)^{-1}
\ee
\item[2.] Proper-time cut-off
\be
R^2[x] = \left\{ \begin{array}{ccc} 1 & : & x \ge 0 \\
erfc( \left| \frac{x}{\Lambda} \right| ) & : & x <0 \end{array}
\right. .
\ee
\end{itemize} 
The Woods-Saxon parametrization is used to approximate a step-function
$ \theta ( x + \Lambda ) $. However, for very small width parameters
$ \sigma $ the iteration scheme fails to converge. Thus for
reasons of numerical stability we choose $ \sigma = 10 MeV $. \\ 
The proper-time cut-off is the standard scheme used in the
NJL soliton model \cite{ARW95,CGGP95}. The corresponding form
of $R$ can be derived by writing the fermion determinant
appearing in the bosonized NJL action as a proper-time integral
\cite{ARW95,CGGP95}. \\
In the iteration scheme we successively solve for the eigenvalues
and eigenvectors of the mean-field Hamiltonian\footnote{In the
notation of part I of this paper the regularized
mean-field Hamiltonian carries a prime which we leave
away in the following} $ H_{MF}^{(i)} $
corresponding to step $i$ of the iteration process.
The matrix to be diagonalized splits up in several
submatrices in the spaces of states with the same 
angular-momentum/parity $(J/P)$ quantum numbers.
Typically we work in a cavity with $R_c=8fm$.  The basis space
usually contains $4000-5000$ states, with an equal number of such
with positive and negative energy. 
The usual iteration scheme, where $ H_{MF}^{(i)} $ is defined
with the eigenvectors of step $i-1$ is found to converge
very slowly. Convergence can be decisively improved by
application of a modified scheme:
\be
H_{MF}^{(i)} = H_{MF} \left( \frac{1}{i-k(i)} \sum_{n=k(i)+1}^{i}
\rho^{(n)} \right)
\ee
with
\be
k(i) = k_0 \; int \left( \frac{i-1}{k_0} \right).
\ee
$ \rho^{(n)} $ is the approximation for the ground-state
density matrix in step $n$. The initial guess $ \rho^{(1)} $
should not be too far away from the self-consistent solution.
The essence of the modified iteration scheme is an averaging
over the density matrices of several iteration steps 
which is reset after $k_0$ steps. For $k_0 = 10$ convergence
is usually achieved after $20-30$ steps. \\
Comparing the results of these calculations with the
solution of the gap equation we find a very good overall
agreement. In Fig.\ \ref{conds} we show the chiral condensate
$ - \langle \bar{\Psi} \Psi \rangle $ as a function of $r$ for
different boundary conditions and cut-off schemes.
The calculations were done with $G_v=0$.
In both cut-off schemes the values for the condensates we
obtain from the gap equation (Woods-Saxon: $1.39 \cdot 10^7 MeV^3$,
proper-time: $1.33 \cdot 10^7 MeV^3 $)  
are reproduced very well. The boundary condition has an effect
on the structure of the vacuum only in the vicinity of the
bag boundary. When the dimension of the basis space is
increased from $4000$ to $6000$, the results are left
unchanged within the line width of the graphical representation. \\
The results show that a translationally
invariant condensate in fact solves the Dyson-Schwinger equation
in the vacuum sector. In the gap equation translational invariance
enters as an a priori assumption which, however, seems to have been 
justified by the results displayed in Fig.\ \ref{conds}. \\
When we switch on the vector potential and solve the constraint
Hartree-Fock problem Eq.\ (\ref{consthf}) we observe that
the chiral condensate slightly increases (i.\ e.\ $\langle \bar{\Psi}
\Psi \rangle $ increases in magnitude), as displayed in Fig.\ \ref{condv}.
For these calculations the proper-time cut-off was chosen and
the constituent-quark mass was fixed to a value $m=600 MeV$. 
If not stated otherwise, we use the nonrelativistic boundary condition
in our calculations.
The coupling constant in the vector channel was adjusted to
the value obtained from the color-current interaction $ G_v= G_s/2 $. \\
Without any vector potential the chiral condensate as obtained
from the gap equation ($1.33 \cdot 10^7 MeV^3 $) is reproduced
with high accuracy even in a smaller cavity with $R_c = 4fm$.
When we increase the cavity radius to $8fm$ the shift of the
condensate effected by the vector potential is less pronounced
than in the small cavity. We thus conclude that the shift will
vanish in the continuum limit. Whereas the net baryon number
can be fixed to zero with the help of a Lagrange multiplier, the
net baryon density will not be a completely smooth function
due to constraints set by the boundary condition.   
The fact that the vector potential couples to the baryon density
offers an explanation for the induced effects which seem to
vanish in the continuum limit. \\ 
\subsubsection{$B=1$-configurations}
\begin{figure}
\vspace{-2cm}
\centerline{\rotate[r]{\epsfig{file=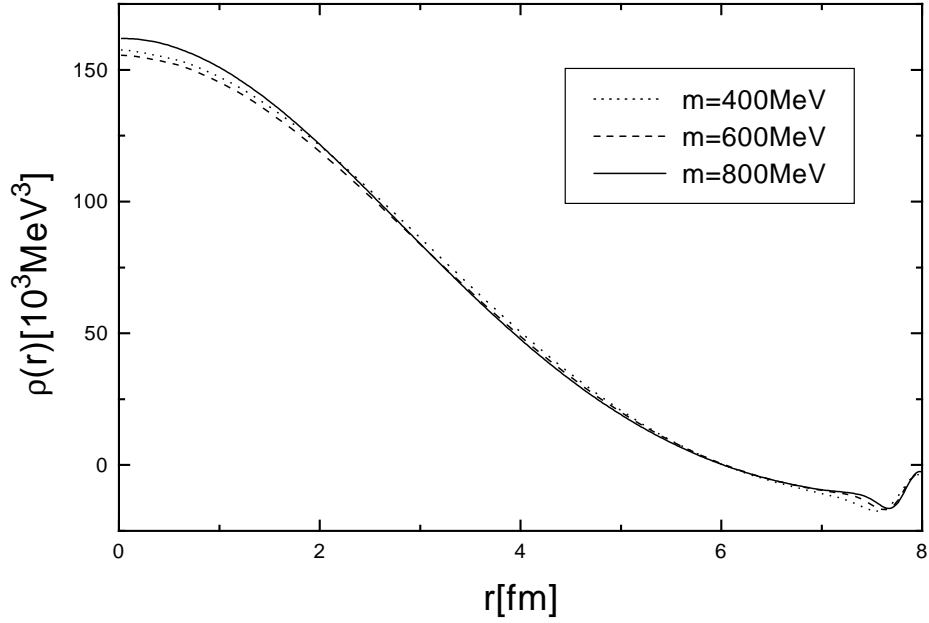,height=14cm}}}
\caption{ $ 3 \langle \Psi^{\dagger} \Psi \rangle $ for different
constituent-quark masses}
\label{3masses}
\end{figure}

\begin{figure}
\vspace{-2cm}
\centerline{\rotate[r]{\epsfig{file=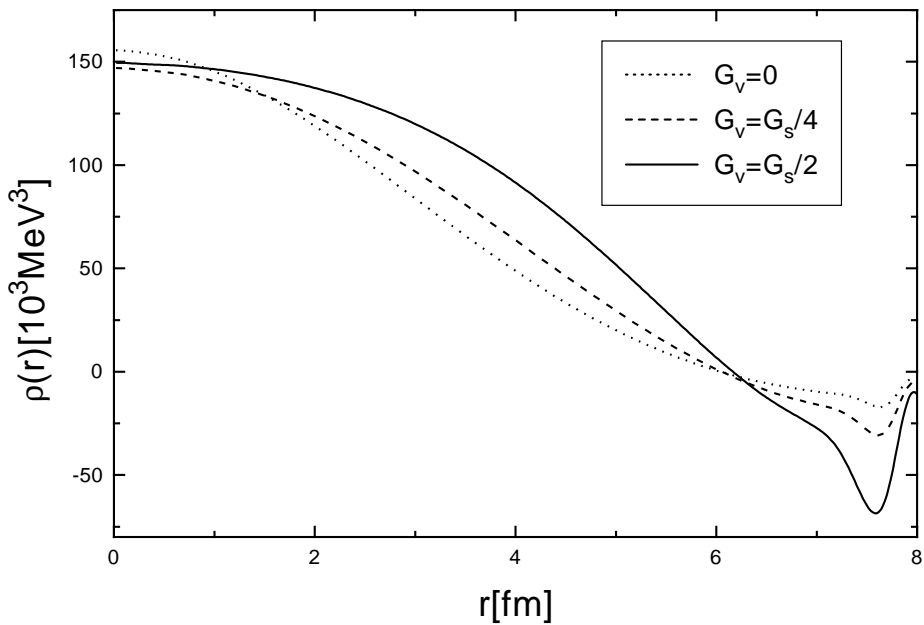,height=14cm}}}
\caption{ $ 3 \langle \Psi^{\dagger} \Psi \rangle $ for different
values of $G_v$ ($m=600 MeV$)}
\label{vec}
\end{figure}

We now turn to self-consistent configurations with unit
baryon number. For the computation of $ B=1$ configurations we
use a similar prescription as for the vacuum, but we
add three valence quarks to the system. This amounts to
replacing the vacuum occupation factors $ \theta (- \epsilon_i) $
by $ \bar{\rho} (i) $ in each expression.
The density matrix $ \bar{\rho} $, which is averaged over
the $ J/T$ projection quantum numbers of the valence quark configuration,
was defined in part I. By such an averaging procedure we
neglect the nonscalar parts of the ground-state density. \\
Properties of the solutions are always extracted from the difference
between the results for the $B=1$ and the $B=0$ system,
which is the vacuum by definition. Effectively, this means that
all additive quantum numbers of the vacuum are renormalized to zero.
Still, vacuum polarization effects are fully accounted for. \\
In such a prescription the baryon number is given by
\be
B = \frac{1}{3} \int d^3 r \left( \sum_{i'} \Psi^{\dagger}_{i'}
\Psi_{i'} R'^2 [\epsilon_{i'} ] \bar{\rho} ( i')
- \sum_i \Psi^{\dagger}_i \Psi_i R^2 [\epsilon_i ] \theta ( -\epsilon_i )
\right).
\label{bnum}
\ee 
The self-consistent single-particle states of the $B=1$ system are
indicated by primed labels.
Note that due to vacuum polarization effects 
the baryon number contribution of the Dirac sea
might be different in the calculations with and without valence quarks.
In order to fix the baryon number difference Eq.\ (\ref{bnum}) to one, 
two slightly different
cut-off functions $R'$ and $R$ were chosen
by readjusting the cut-off scale $ \Lambda $.
The difference of the original and the readjusted cut-off scale
was always found to be less than $1 MeV$.
If the same cut-off is used in the two calculations, the
resulting field configurations look very much the same as
the configurations with readjusted cut-off, but the total energy
of the system including valence quarks is found to be lower
than the vacuum energy. The energy is defined as the expectation
value of the NJL Hamiltonian in a ground state with regularized 
one-body density
\be
E= \sum_{i'} \epsilon_{i'} \bar{\rho} (i') R'^2 [\epsilon_{i'}]
- \frac{1}{2} \sum_{i'j'} \langle i'j' | \bar{v}_{NJL} | i'j' \rangle
\bar{\rho} (i') \bar{\rho} (j')
R'^2 [\epsilon_{i'}] R'^2 [\epsilon_{j'} ],
\ee 
where the two-body interaction is derived from the NJL Lagrangian
Eq.\ (\ref{fierz}) and given by
\be
\langle i j | v_{NJL} | i j  \rangle = G \int d^3 r \; \bar{\Psi}_i
\gamma_{\mu} \frac{\lambda_a}{2} \Psi_i \; \bar{\Psi}_j \gamma^{\mu}
\frac{\lambda_a}{2} \Psi_j .
\ee 
When we include the vector part of the self energy, we subtract
the baryon density as obtained from the vacuum calculation 
and define the remaining contribution as the effective
baryon density entering the equations as a source term for the
vector potential. In the continuum limit, where boundary effects
become unimportant, this corresponds to the introduction of
a Lagrange multiplier, similar to Eq.\ (\ref{consthf}). 
In all the calculations for the $B=1$ system the proper-time cut-off
was used. \\
In Fig.\ \ref{3masses} the baryon density is displayed for
different constituent-quark masses and $G_v=0$. The coupling
constant in the scalar channel was fixed to reproduce
phenomenologically reasonable values for the chiral
condensate of the order of $ (-240 MeV)^3 \approx 1.4 \cdot 10^7
MeV^3 $. 
The self-consistent field configurations are spread out
over the entire cavity. The structure of the solutions
changes very little when the constituent-quark mass is varied.
Obviously the scalar field alone does not
provide enough attraction to form a bound, solitonic object. \\
From the NJL soliton model it is known that below a critical
constituent-quark mass no solitons are found \cite{ARW95}.
Since the density profile is invariant
under variations of the constituent-quark mass in the region
below $1 GeV$,
we do not expect that such a phase transition occurs
in our model for any constituent-quark masses up to
a few $GeV$. \\
Due to vacuum polarization effects the baryon density becomes
negative in the vicinity of the bag boundary. 
The masses of the configurations are $ 3769 MeV $,
$ 4586 MeV $ and $ 5440 MeV $ for $ m $ equal to $ 400 MeV, 600 MeV $
and $ 800 MeV$, respectively.
The large values for the masses indicate that the vacuum gets
considerably polarized in the presence of valence quarks. \\
In order to demonstrate the large spatial extension of the
density profiles more clearly and to discuss finite-size
effects in the $B=1$ sector we have solved the mean-field
equations for different values of the basis parameter $R_c$.
For this purpose the constituent-quark mass was chosen as 
$ m=600 MeV $. 
\begin{figure}
\vspace{-2cm}
\centerline{\rotate[r]{\epsfig{file=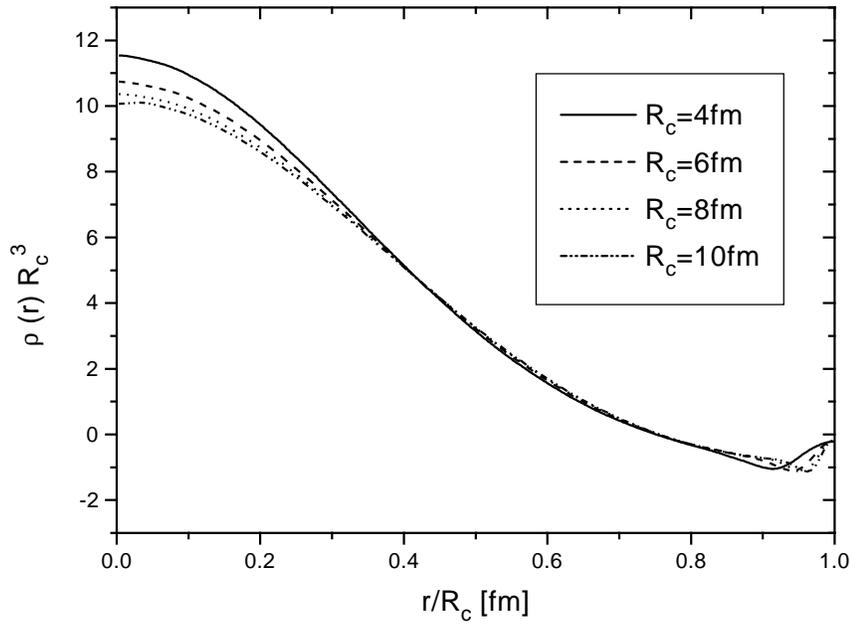,height=14cm}}}
\caption{ $ 3 \langle \Psi^{\dagger} \Psi \rangle R^3_c $ 
for different values of $R_c$ ($m=600MeV $).
The factor $ R^3_c $ accounts for the normalization condition
and was included in order to compare the different profiles.}
\label{4r}
\end{figure}
In Fig.\ \ref{4r} the corresponding baryon density is shown
as a function of $r/R_c$. After multiplication with a scale
factor $ R^3_c $ the profiles look very similar, thus indicating
the basis dependence and spatial extension of the solutions.
In terms of the total energy of the configurations, however, a radius
of $R_c = 8fm $ seems sufficient to reproduce the continuum limit
with reasonable accuracy. This is indicated by the functional dependence
of the energy on $R_c $ which is displayed in Table \ref{diffr}. \\
\begin{table}
\begin{center}
\begin{tabular}{|c|c|c|c|c|}
\hline
$R_c [fm]$ & $4$ & $6$ & $8$ & $10$ \\
\hline
$E_{tot} [MeV]$ & $4454$ & $4542$ & $4586$ & $4587$ \\
\hline
\end{tabular}
\caption{The total energy of the $B=1$ configuration
as a function of the basis parameter $R_c$ ($m=600 MeV$).}
\label{diffr}
\end{center} 
\end{table}
It is helpful to map the degrees of freedom in our approach
to those used in the NJL soliton model. The simplest version of the
latter uses a scalar-isoscalar field $ \sigma $ and a
pseudoscalar-isovector field $ \mbox{\boldmath$\pi$\unboldmath} $.
In terms of mode sums as in Eqs.\ \ref{scalar} and \ref{vector}
these fields are given by
\begin{eqnarray}
\sigma & = & \sum_i \bar{\Psi}_i \Psi_i \; R^2 [\epsilon_i] \; \bar{\rho} (i) 
\nonumber \\
\mbox{\boldmath$\pi$\unboldmath} & = & \sum_i \bar{\Psi}_i \gamma_5
\mbox{\boldmath$\tau$\unboldmath} \Psi_i \; R^2 [\epsilon_i] 
\; \bar{\rho} (i) .  
\end{eqnarray} 
By construction of $ \bar{\rho} $, we obtain 
$ \mbox{\boldmath$\pi$\unboldmath} =0 $.
In the NJL soliton model the fields are usually constrained
to the chiral circle $ \sigma^2 ( { \bf r} ) +
\mbox{\boldmath$\pi$\unboldmath}^2 ( {\bf r} ) = f_{\pi}^2 $,
where $ f_{\pi} $ is the pion decay constant. 
When this constraint is dropped the solitonic solutions
have been found to become unstable and the $ \sigma $ field
develops a sharp structure localized around the origin \cite{SMGG92,WT92}.  
However, it was then shown in \cite{SWRA93} that the
collapse can be prevented by 
properly fixing the baryon number. \\
Clearly such results are relevant in the present context,
since by varying the $ \sigma $ field and constraining
the pion field to zero we leave the chiral circle.
In all the calculations performed we have never
observed a collapse of the field configuration.
Even when the baryon-number constraint is dropped
a localized field configuration does not develop.
Note, however, that the prescription for fixing the baryon
number is different from the one used in Ref.\ \cite{SWRA93}. \\
Based on the results obtained by Sieber et al.\ \cite{SMGG92} 
a simple explanation for this observation can be given.
They have found that the collapse of the soliton is related
to the valence level being lowered to negative energies, whereas
the rest of the spectrum remains basically unchanged.
A necessary condition
for such a phenomenon to occur is the violation
of charge conjugation invariance. In \cite{SMGG92} 
charge conjugation invariance is broken due to a nonvanishing
pion field.
However, the mean-field Hamiltonian we use does not
contain pionic degrees of freedom. In addition to that, the model
space we have defined in order to numerically solve the 
Dirac-Hartree-Fock problem is $C$-invariant. As a result,
the symmetry of the single-particle spectrum
under inversion of the energy axis is reproduced with
high accuracy and a collapse of the fields driven
by a level plunging into the Dirac sea can never occur. \\
From the results for the $B=1$ sector we conclude that the pseudoscalar contribution
to the self-energy is of major importance for the emergence
of the soliton in the grand-spin approach. In our model the pseudoscalar field is
constrained to vanish by reasons of symmetry. In the NJL soliton model
the self-consistent solutions exhibit a sizeable, surface-peaked
pion field \cite{MG91}, which in a bosonized formulation takes the role
of the pseudoscalar mean field. \\
In order to quantify the importance of pionic contributions
we have estimated the relative size of the different terms of Eq.\ (\ref{fierz}) in the ground state of the nucleon.
The pionic contributions vanish by averaging the density
matrix of the system, so the idea is to evaluate the corresponding
matrix elements before taking the average. We thus consider a state
vector with three valence quarks coupled to
$ (J^p,T) = (\frac{1}{2}^+, \frac{1}{2}) $, as discussed in
part I of the paper.
The valence wave function is taken from a typical soliton
configuration in the NJL model, where only the $ \sigma $
field is taken into account. The latter restriction is necessary
in order to extract single-particle states with the desired
symmetry properties. By means of such a construction we try
to mimic a hypothetic solitonic solution which is consistent
with the specific assumptions our work is based on. \\ 
More specifically, we use a Woods-Saxon
parametrization of the selfconsistent field obtained
by Meissner et al. \cite{MG91} with the chiral constraint
and a constituent-quark mass of $m=600 MeV$.
Such a field configuration should be a fairly good representative
for soliton solutions which are dominated by the valence quarks.
For simplicity we assume the absence of vacuum polarization
and, in the spirit of an estimate, determine the properties of the system by the valence configuration alone. \\
The interaction defined in Eq.\ (\ref{fierz}) picks up contributions of different Lorentz and isospin structure.
In order to estimate the importance of the different terms
for the binding of the soliton we have computed the expectation value of the corresponding contribution
to the interaction Hamiltonian. The coupling
constants $ G_s $ and $ G_v$ can in general 
be regarded as independent parameters. \\ 
As an important
result we find that the net attraction is provided
by the scalar-pseu\-do\-sca\-lar channel, whereas the sum of
the vector-axialvector terms gives a positive
contribution to the total energy. The scalar-isoscalar
and pseudoscalar-isovector terms are both attractive
and for the chosen ground-state configuration we
find
\be
\frac{ \langle N | \int d^3 r 
( \bar{\Psi} \gamma_5 \mbox{\boldmath $\tau$ \unboldmath} \Psi)^2 | N \rangle }{ - \langle N | \int d^3 r ( \bar{\Psi} \Psi )^2
| N \rangle } \approx 2.4 . 
\ee 
The pionic term is thus expected to add a contribution
to the binding of a hypothetic soliton at least as important
as the scalar-isoscalar term.  
When the constituent-quark mass and thus the depth of the potential
well is varied between $ 400 $ and $ 800 MeV $ the ratio of the
different contributions stay in the same order of magnitude
and the above statements on the sign of the terms remain valid. \\
It can thus be concluded that our results for the $B=1$ sector
are in accordance with such a simple estimate on the binding
effects due to a local, chiral interaction. Our numerical investigations
show that neglecting the attractive pionic component leads to
a drastic change in the structure of the ground state. In view
of a realistic description of the nucleon within
the NJL model the hedgehog ansatz, allowing for a
nonvanishing pion field, is clearly superior to our treatment of
symmetries. \\
However, a few aspects of the approach have not yet been illuminated.
It has been pointed out that topologically nontrivial solitonic 
field configurations should be represented in a model space
of states that satisfy appropriate boundary conditions \cite{ARSW94}.
The basis space suited for, e.\ g., scalar and pseudoscalar
fields is different from the appropriate space for vector
and axial-vector fields. Topological features
of our field configurations cannot be discussed in analogy
to the NJL soliton model, where topologically distinct classes of fields
are characterized by different winding numbers of the chiral angle.
Since the pion field is missing, we do not have the same
classification of topological structures as in the NJL soliton model.
Still we want to examine the
influence of the boundary condition imposed on the basis states
on the structure of the self-consistent solutions.
\begin{figure}
\vspace{-2cm}
\centerline{\rotate[r]{\epsfig{file=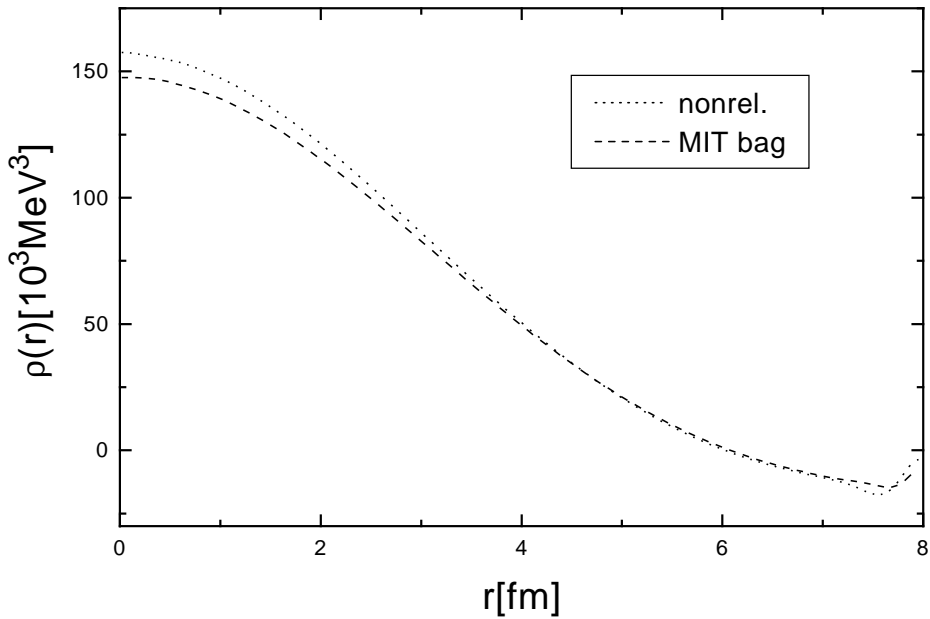,height=14cm}}}
\caption{ $ 3 \langle \Psi^{\dagger} \Psi \rangle $ as obtained
in model spaces with different boundary conditions.
The constituent-quark mass was fixed to $m=400 MeV$.}
\label{2bound}
\end{figure}
In Fig.\ \ref{2bound} the baryon density as obtained with the
nonrelativistic and the MIT-bag boundary condition is displayed.
The two density profiles are almost identical.
The energy obtained with the MIT-bag boundary condition is slightly
higher ($3855 MeV$ vs.\ $3769 MeV$).
As already found in the vacuum
sector, the boundary condition only has a minor
influence on the structure of the solutions. \\
When we switch on the vector potential, the
density profile becomes broader (Fig.\ \ref{vec}).
This is expected from the repulsive character of the vector interaction
in the particle-particle channel. The energy of the configurations
is increased compared to the value obtained with $G_v=0$.
When $G_v$ exceeds a critical value $ G_v^{crit} \approx G_s $
the iteration scheme fails to converge. 
An interesting situation occurs when the sign of the vector interaction is reversed.
Clearly such a situation is unphysical, since the vacuum
becomes unstable for a strongly repulsive $ q \bar{q} $
interaction. However, before the instability sets in a localized, solitonic
system is formed. While lacking any physical relevance the formation
of a bound state can be viewed as a test for the numerical implementation
of the Dirac-Hartree-Fock approach. It thus becomes evident that
the basis space contains the necessary high-frequency components
to resolve a well-localized object. \\
With no other degrees of freedom besides the scalar and the
vector channel accessible, we conclude that the NJL model does not
possess solitonic solutions in the $B=1$ sector if a
rotationally and isospin invariant mean-field
configuration is used. It is crucial for the emergence of the soliton
to truncate the symmetry group to an invariance under simultaneous
rotations in isospin and coordinate space, as inherent in
a grand spin zero configuration. \\
The results indicate that the prescription to obtain
an invariant ground-state density by means of spin-flavor 
averaged density matrices is insufficient
in the context of the NJL model. This conclusion, however, might
loose its validity within a more general class of models. The 
phenomenological success of the MIT-bag model as well as the
nonrelativistic quark model \cite{FH68} indicates that confining
forces play a crucial role for the binding of baryons.
The confining forces were completely neglected so far,
because the NJL model allows free quarks as asymptotic states. 
A major goal for the further development of chiral quark
models is therefore the incorporation
of a self-consistent description for confinement. 
\subsection{Modeling Confinement}
The ground-state configurations obtained for the $B=1$
system are not suited for the phenomenological investigation
of the nucleon, since on mean-field level the valence quarks are not
bound. \\ 
Our main interest is the excitation spectrum of nucleon
rather than the ground state. In an RPA calculation the information
about ground-state structure enters via the single-particle energies
and wavefunctions. We therefore need to construct a model for
the nucleon from which we extract the necessary information. \\
The situation is similar as in the early days of RPA
calculations for low-lying excitations of nuclei. There a phenomenological
single-particle basis was used together with a parametrization
for the residual interaction \cite{GGS66,MMG67}. Despite the
lack of consistency excitation energies and transition strengths
could be reproduced quite successfully with such an approach. \\
As a prototype of a simple model for the nucleon we consider
the MIT-bag model \cite{CJJTW74}, 
where the determination of a set of positive-
and negative-energy eigenstates reduces to the task of solving
a nonlinear equation. However, for the examination of the
dynamical response of the nucleon this model seems to be too crude.
According to the tensor-RPA description
the RPA-excited states might have contributions from
monopole modes admixed whenever this is allowed by angular-momentum
selection rules.
The monopole modes are described by $ph$ pairs coupled
to $J^P =0^+$ and can be viewed as radial oscillations of the
ground-state density. 
We expect that in the MIT-bag model
such modes cannot be adequately described due to the rigid
bag boundary. The results to be discussed in the
next section support the idea that an MIT-bag potential 
does not provide a suitable ground state for RPA calculations. \\
In order to account for surface vibrations and monopole modes
the MIT bag has to be replaced by a smoother potential. Let us
consider scalar potentials of the form
\be
m(r) = c r^{\alpha},
\ee
where the MIT bag is recovered in the limit $ \alpha \rightarrow
\infty $. The excitation spectrum of the system without any
$ph$ interaction can be obtained from the single-particle spectrum
of the Dirac Hamiltonian containing the scalar self-energy
$ m(r) $. These states will get shifted when the RPA
residual interaction is turned on. It is well known that due to
ground-state correlations one or even a couple of these levels will
get rapidly lowered (see, e. g., \cite{SE73}). Hence we
expect an enhanced level density at low energies. A linear
potential $ (\alpha=1) $ will lead to a high density of excited
states already on mean-field level and seems problematic for
that reason. We therefore decided to choose an 
harmonic-oscillator potential
$ (\alpha=2) $. \\
Anticipating some of the results of the next chapter, for
a standard nucleon configuration with a root-mean-square radius (rms)
of about $ 0.8 fm $ the problem of an overabundance of
low-lying states cannot be solved by simply picking a specific
$ \alpha $. In order to obtain a fair description of the
baryon spectrum we are forced to consider a quark core of
drastically reduced size. It is generally assumed that the experimental value for the root-mean-square radius contains
sizeable contributions from the $ q \bar{q} $ polarization
cloud of the nucleon, the quark core being considerably smaller.
Obtaining reliable experimental data for the size of the
quark core is extremely difficult. However, a few attempts
in that direction have been made, with estimates ranging from
$ 0.2 fm $ \cite{Ore78} to $ 0.5-0.6 fm $ \cite{BRW86}. \\ 
In our approach the quark core is modeled by the mean-field configuration, whereas the $ q \bar{q} $ polarization cloud is accounted for
by RPA ground-state correlations. Thus,
besides a standard MIT-bag configuration, we consider
an oscillator potential giving a nucleon rms of about $ 0.3 fm $.
In table \ref{pots} we display the characteristics of the
different models for the ground state. The calculations presented
in the next chapter are based on these two alternative descriptions. \\ 
\begin{table}
\begin{center}
\begin{tabular}{|c|c|c|c|}
\hline
Potential & $ \alpha $ & $ \langle r^2_E \rangle_p^{
\frac{1}{2}} [fm] $ & $ E_{val} [MeV] $ \\
\hline \hline
MIT-bag & $\infty$ & 0.71 & 405 \\ \hline
oscillator & 2 & 0.33 & 1052 \\ \hline
\end{tabular}
\caption{Root-mean-square charge radius and valence-quark
energy for the different parametrizations of the potential}
\label{pots}
\end{center} 
\end{table}
For the harmonic-oscillator potential the Dirac equation for the
central-force problem was solved numerically by a Runge-Kutta method
with adaptive stepsize control. In case of the MIT-bag potential
only one nonlinear equation for the eigenvalues had to be
solved. In both cases the current-quark mass was fixed to
$ m_0 = 5 MeV $. The numerical method for the first order
differential equation has been checked by comparison with the
results for an MIT-bag potential and a scalar potential of
the form $ m(r) = c r $ \cite{C76} which both can be obtained
semi-analytically. \\  
In the simple potential picture there is nothing
that makes a valence-quark configuration with the quantum
numbers of the nucleon energetically preferable to
a $ \Delta (1232) $ configuration. For this reason the
ground-state energy should be considered as the $ N-\Delta $
average. Due to the uncertainty principle 
we obtain a quite large value for the valence 
energy in case of the harmonic-oscillator
potential. This might be an indication that the 
total energy of the ground state
which picks up a contribution from the single-particle
energies is overestimated in this model.  
Note, however, that
in a RPA description the energy is lowered by ground-state
correlations \cite{RS80}. Such correlations account for the
deviation of the physical state vector from a simple, antisymmetrized
many-body wave function (e. g. a Slater determinant) and always
decrease the energy of the ground state. The concept is meaningful
even in the context of bag models which cannot be derived as
self-consistent solutions of a field theory. In that regard
especially the problem of restoration of broken symmetries
has been studied. It has been found that, e. g., center-of-mass
corrections lead to a significant decrease of the energy of bag states
\cite{DGHW83}. In the NJL soliton model ground-state correlations
have been found to lower the energy of the mean-field configuration
by a couple of hundred $MeV$ \cite{ARW95}. 
However, the RPA is only expected to work well when 
ground-state correlations are not too sizeable.
Otherwise an extended scheme as in Ref.\ \cite{SE73} has to
be taken into account which incorporates
dynamical correlations also in an explicit fashion. 
This issue will be discussed
in more depth in the next chapter. \\
An effective way to model ground-state correlations 
would be to consider a small-size MIT-bag configuration
with modified boundary conditions on the bag surface,
the so called ``Little Bag" \cite{BR79}.
The modified boundary conditions are due to the chiral
fields outside the bag which, in a complete RPA treatment,
are described by $ q \bar{q} $ admixtures to the ground state.
It can be shown that the chiral fields always lower the energy
of the valence level \cite{Mul84}, thus supporting the arguments
given above. \\
In a much simpler fashion we could shift the single-particle
levels without even changing the wavefunctions by introducing
a (constant) time-like component of a vector potential.
This shows that within the scope of the simple models considered
here the valence energy is not a particularly meaningful quantity.
It is rather the single-particle excitation energies which
are relevant in an RPA context. The latter are clearly insensitive
to a shift of the energy scale. \\   
Our choice of a central mean-field potential is compatible with
the symmetries we demand for the eigenstates. 
Besides the restrictions for rotational and isospin symmetry set
by the tensor-RPA method we require parity eigenstates.
These requirements can as well be fulfilled by a potential
transforming as the time-like component of a Lorentz vector.
However, the inclusion of a vector potential in the Dirac equation 
might lead to solutions which are not normalizable. Recently
it was shown that the wavefunctions obtained in a linear
vector potential acquire a non-integrable singularity,
whereas the solutions in the corresponding scalar potential
are well behaved \cite{S95}.
In order to avoid such problems and to keep the model for the ground-state
as simple as possible we restrict ourselves to scalar potentials. 
The eigenenergies and eigenfunctions obtained with the
above parametrizations serve as an input for the tensor-RPA
method. In the next section we discuss a first application
of this method to the excitation spectrum of the nucleon. 
\section{The Excited States}
\begin{figure}
\vspace{-1cm}
\centerline{\epsfig{file=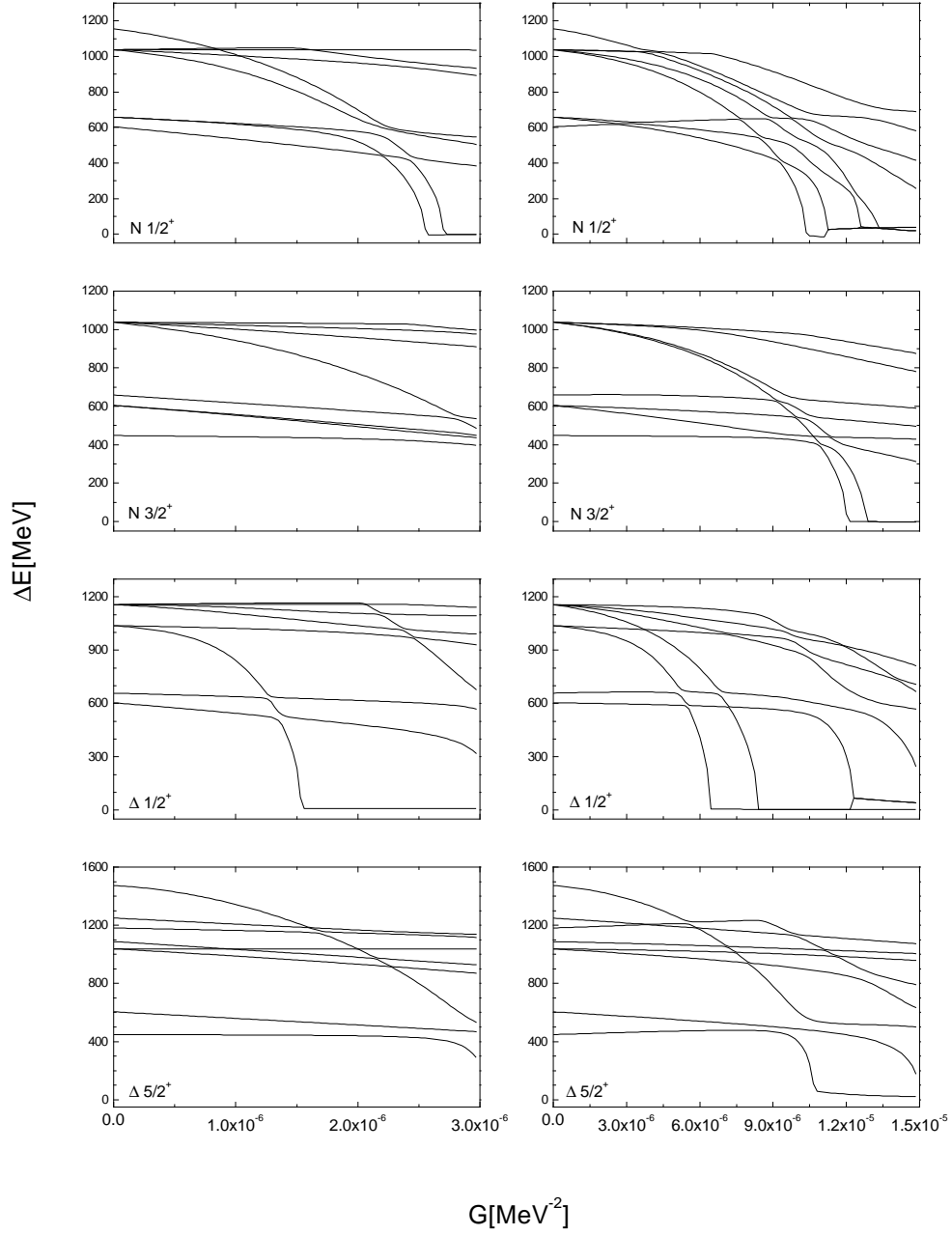,height=22cm}}
\vspace{-2cm}
\caption{The excitation energies as a function of $G$ for the
MIT-bag mean field. 
Left column: $ \sigma -
\mbox{ \boldmath $ \pi $ \unboldmath } $ interaction.
Right column: Color-current interaction.}
\label{usualmit}
\end{figure}

\begin{figure}
\centerline{\epsfig{file=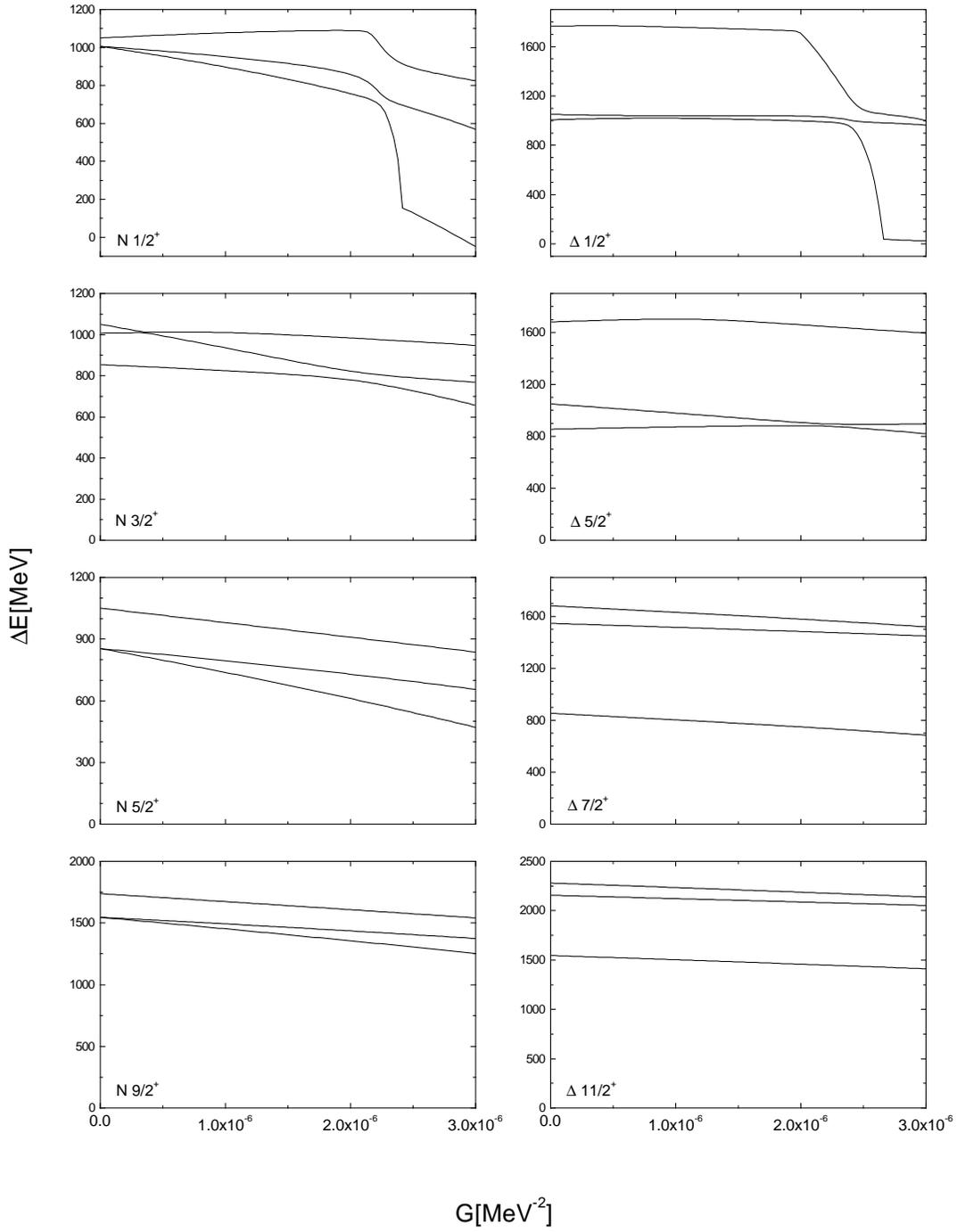,height=23cm}}
\vspace{-3.5cm}
\caption{The excitation energies as a function of $G$ for the
harmonic-oscillator mean field in combination with the color-current
interaction.}
\label{small}
\end{figure}

\begin{figure}
\vspace{-1cm}
\centerline{\epsfig{file=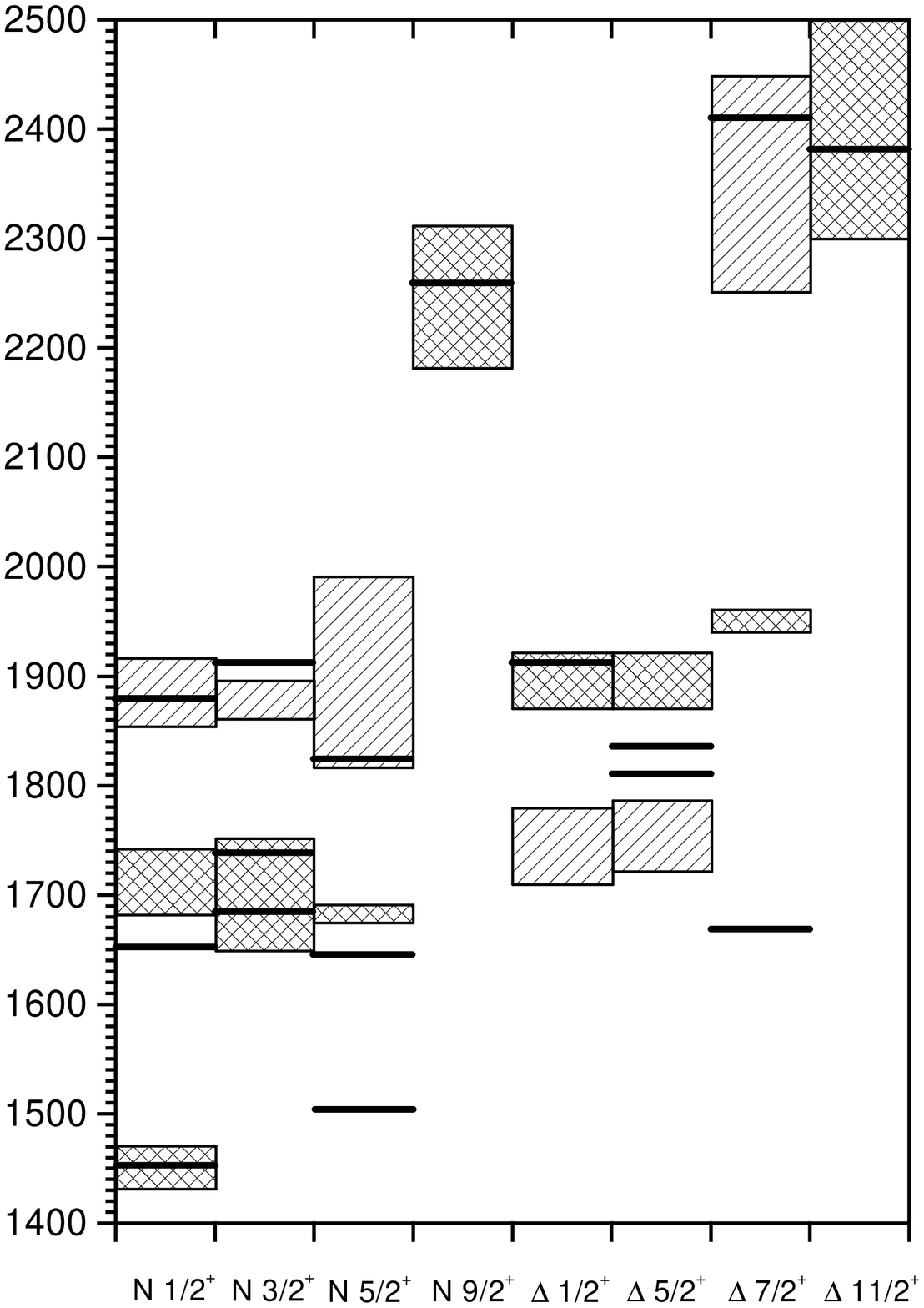,height=23cm}}
\vspace{-2cm}
\caption{The spectrum of positive-parity baryons
obtained with the color-current interaction at
$ G = 2.35 \cdot 10^{-6} MeV^{-2} $. RPA results are indicated by
bars and experimental data are represented by hatched areas.} 
\label{specfit}
\end{figure}

Although in the last 20 years a variety of models for
baryons have been proposed \cite{Bha88}, we are still far from
a complete understanding of the baryon spectrum. Experimentally
about 20 resonances below $2 GeV$ are confirmed in the non-strange sector. \\
The phenomenologically most successful models 
have their origin in the nonrelativistic quark model \cite{FH68}
which was systematically applied to the problem of baryon
spectroscopy by Isgur and Karl \cite{IK78}.
Initially, such formulations contained
three constituent quarks which interact via a harmonic-oscillator
potential and a residual interaction of spin-color type which 
is treated perturbatively. With such an ansatz the baryon spectrum
in the negative-parity channel could be satisfactorily reproduced,
whereas with the same parameter set no acceptable description
of the positive-parity data was obtained. \\
The ``parity problem" was overcome only recently when the formulation
of the constituent-quark model was modified.
On the level of perturbation theory a satisfactory description
of the data in both parity channels is obtained when the
spin-color interaction of Isgur and Karl is replaced by
a spin-flavor interaction \cite{GR95}.
A nonperturbative treatment of a generalized two-body interaction
in a model space leads to a good overall agreement 
with the data as well \cite{BIL94}. 
Despite these successes the constituent-quark models leave
a number of questions unresolved. \\
First of all it still remains difficult to establish a link to QCD, since
the elementary degrees of freedom are constituent quarks which
are quite complicated objects in terms of the current quarks of QCD
\cite{VLKW90}. \\ 
Besides this conceptual problem there are two
issues to be clarified. As a result of most formulations of
the constituent-quark model unobserved states occur in the
spectrum. It needs to be clarified if these states exist in nature
and have not yet been detected experimentally or if they are
incorrectly predicted by the constituent-quark model.
The problem of unobserved states seems to be quite general
and is also encountered in attempts to describe the baryon spectrum
in the MIT-bag model even for relatively small model spaces
\cite{DJ76,S78}. \\
The other problem that needs further inspection is related to
vacuum structure. The mass scale of baryon resonances predicted
by the constituent-quark model is such that sizeable admixtures
of $ q \bar{q} $ states are expected even for quark masses
of the order of $ 200-300 MeV$. Such contributions
are completely neglected in the constituent-quark model.
The assumption that the spectral functions of QCD are dominated
by the lowest excitations of the QCD vacuum leads to
a low-energy effective theory known as chiral perturbation
theory \cite{L94}. The phenomenological success of
chiral perturbation theory indicates that such an assumption
seems well justified \cite{DGH92}.   
Thus mesonic admixtures to the baryon wavefunctions
are easily conceivable.  
The Skyrme model \cite{Sky61} constitutes an extreme 
but quite successful description
of baryons in terms of meson fields only. \\
The tensor-RPA method as developed in part I of this paper 
allows to treat the valence excitations and the excitations
of the Dirac sea in the same framework. It may serve as a bridge
from constituent-quark models to chiral models. \\
The tensor-RPA matrix elements split up into a mean-field 
and a residual interaction part. The mean-field part is
defined by single-particle energies $ \epsilon_k $ which
we take from the solution of the Dirac equation for
the confining potentials discussed in the previous section. \\
For the residual interaction part we need to specify a particle-hole
($ph$) interaction. A simple choice that has proven to be
applicable in the context of a RPA description of mesons
\cite{KLVW90} is the color-current interaction Eq.\ (\ref{colcu}).
As a second model we use an NJL-type interaction in the $ \sigma $-
and $ \mbox{ \boldmath $ \pi $ \unboldmath } $-channel
\be
{\cal L}_{int} = G \left[ ( \bar{\Psi} \Psi )^2 - ( \bar{\Psi} \gamma_5
\mbox{\boldmath $ \tau $ \unboldmath } \Psi )^2 \right]. 
\label{NJL}
\ee
The Lagrangian defined above is invariant under 
$ SU(2)_L \times SU(2)_R $ chiral transformations. Recently
it has been claimed by Glozman and Riska that a 
pseudoscalar meson exchange between constituent quarks 
could explain the structure of the baryon spectrum \cite{GR95}.
With the interaction Eq. (\ref{NJL}) we have effectively introduced
a local pion and sigma exchange between current quarks.
Such a Lagrangian can be extracted from the Sigma Model
when the meson masses are going to infinity \cite{Bha88}.  
The arguments
in favor of a pseudoscalar meson exchange interaction given in Ref.\
\cite{GR95} were based on the group-theoretical structure of
the nonrelativistic reduction of their model rather than on
the detailed form of such an interaction. Our goal is to study
the effects of pseudoscalar meson exchange on current-quark level
with the above Lagrangian. \\
A chirally invariant point-like interaction between quarks
has been motivated from various approaches describing the
QCD vacuum \cite{AR92,D96}. All these approaches neglect
the backreaction of the gluon fields to quark sources and
only describe quark interactions in the background of
the nonperturbative gluon vacuum. In the extreme case of an
MIT-bag ground state the quarks expel the gluon condensate
from a spherical cavity leaving the perturbative vacuum inside.
In such a scenario the use of a perturbative one-gluon 
exchange interaction seems more natural than a NJL interaction.
A finite-range interaction would lead to retardation effects
which are not accounted for in the present RPA formulation.
A simple estimate shows that for the range of excitation energies
we are interested in retardation effects might play an
important role. Therefore we consider a local color-current
interaction as an equally suitable description as a finite-range
gluon exchange without retardation. \\
The combination of one of the different mean-field configurations
discussed in the previous section with a $ph$ interaction
of local color current or local $ \sigma -
\mbox{\boldmath $ \pi $ \unboldmath } $ type clearly does not
constitute a consistent, symmetry conserving Dirac-RPA scheme.
Hence we expect spurious admixtures in the excitation spectrum.
However, such admixtures can be excluded by restricting
the calculation to excited states with positive parity.
In the small amplitude limit a state generated by acting
with a symmetry transformation on a positive-parity ground 
state carries the same parity quantum numbers as the generators
of the symmetry group
\begin{eqnarray}
exp( i \theta_a S_a) \; | N \rangle & = & | N \rangle + | \Delta N \rangle
\nonumber \\
P \; | \Delta N \rangle & \approx & i \theta_a P S_a P \; | N \rangle .
\end{eqnarray}
In these expressions $ S_a $ denotes the generators of the
Lie-group under consideration and $ P $ is the parity operator.
In our model for the nucleon the relevant symmetries which are
broken in the ground state are translational and chiral
invariance\footnote{Due to the small current-quark masses chiral
symmetry is also broken explicitly. Our discussion applies to the
case of vanishing current-quark masses}.
Chiral symmetry is broken by the scalar mean-field potential.
The generators of these groups are both negative-parity operators,
so that in the positive-parity channels the RPA states are orthogonal to
the spurious modes. \\
Since the ground state is modelled by solutions of the Dirac equation
we are in a position to account for $ q \bar{q} $ components
in the wavefunctions of the excited states. Note, however, that
the Dirac sea in such a description is given by localized negative-energy
states. Thus our ansatz includes those mesonic excitations that
are strongly correlated with the quark core of the nucleon. 
It would be desirable to find a RPA description
for the whole spectrum of mesonic excitations including
the pseudoscalar Goldstone modes of the QCD vacuum which are decoupled
from the regions of the Dirac sea polarized by the quark core.
However, such a general approach within a confining
quark model seems extremely difficult
and is left for future investigations. \\
For the numerical realization of the tensor-RPA scheme we compute
the RPA matrix $M$ and the metric tensor $N$ as defined in
part I for various combinations of the mean-field
potentials and the $ph$ interactions discussed above. 
In a compact notation, the RPA-equation is given by \cite{HGLM96}
\be
\left( \begin{array}{cc}   M^{(1)} & -M^{(2)}  \\
-M^{(3)} & M^{(4)} \end{array} \right) \left( \begin{array}{c}
X \\ Y \end{array} \right) = \omega_{\Delta}
\left( \begin{array}{cc}  N^{(1)} & -N^{(2)} \\
-N^{(3)} & N^{(4)} \end{array} \right) \left( \begin{array}{c}
X \\ Y \end{array} \right) .
\label{tensorrpa}
\ee
with
\begin{eqnarray}
M_{(mi)k,(nj)l}^{(1)} & = & \sum_{ \Gamma} C_{k,l}^{\Gamma}
\langle \Gamma_N || [ A_{mi} ( \bar{\Gamma}_k ) , H , A_{nj}^{\dagger}
(\Gamma_l ) ]^{\Gamma} || \Gamma_N \rangle \nonumber \\
M_{(mi)k,(nj)l}^{(2)} & = & \sum_{ \Gamma} C_{k,l}^{\Gamma}
\langle \Gamma_N || [ A_{mi} ( \bar{\Gamma}_k ) , H , A_{nj}
(\bar{\Gamma}_l ) ]^{\Gamma} || \Gamma_N \rangle \nonumber \\
M_{(mi)k,(nj)l}^{(3)} & = & \sum_{ \Gamma} C_{k,l}^{\Gamma}
\langle \Gamma_N || [ A_{mi}^{\dagger} ( \Gamma_k ) , H , A_{nj}^{\dagger}
(\Gamma_l ) ]^{\Gamma} || \Gamma_N \rangle \nonumber \\
M_{(mi)k,(nj)l}^{(4)} & = & \sum_{ \Gamma} C_{k,l}^{\Gamma}
\langle \Gamma_N || [ A_{mi}^{\dagger} ( \Gamma_k ) , H , A_{nj}
(\bar{\Gamma}_l ) ]^{\Gamma} || \Gamma_N \rangle ,
\label{rpamatr}
\end{eqnarray}
where $ H $ is the full Hamlitonian and 
$ C_{k,l}^{\Gamma} $ are recoupling coefficients.
The metric tensor $N$ is obtained from eq.\ (\ref{rpamatr}) 
when the Hamiltonian is omitted and the double commutators
are replaced by ordinary commutators.
For the computations presented here we use 
the exact expressions for the many-body matrix elements
of Eq.\ (\ref{rpamatr}) as given in part I. \\
In order to check the rules for the reduction of the overcomplete
$ph$ space we have computed the metric tensor in the full
space of $ph$ states. It was found that whenever a set of
energetically degenerate states was linearly dependent according to
the reduction scheme the metric tensor $N$ was singular in
that subspace. When the full space of dimension $d$ was reduced
to the smaller dimension $d_r$, the determinant taken in the
reduced space became nonzero. \\
The results for the excitation spectrum presented here were obtained
with a $ph$ basis of dimension $50$. With a larger model space 
the energies of the lowest excitations are shifted.
However, this shift can be compensated by rescaling the coupling
constant. At the rescaled value of $G$ the spectrum looks very
similar to that which was obtained with the standard basis of
dimension $50$. This observation seems to indicate that the
dynamics of the system depends only on
a certain combination of the coupling constant and
the cut-off in $ ph $ space. A similar
behaviour is found in the vacuum sector of the NJL-model,
where the relevant parameter is $ G \Lambda^2 $ \cite{VW91}. 
Note, however, that in the latter case $ \Lambda $ defines
a cut-off in the single-particle energy. \\  
For any excited state with fixed quantum numbers $(J^p,T)$ the 
model space is spanned by the $ph$ states with lowest energy
with respect to the mean-field potential. Valence excitations
and excitations of the Dirac sea contribute in an equal
manner. The $ph$ states are chosen according
to angular momentum/isospin/parity selection rules, including
any $ph$ combination which can be coupled with the
ground state to the desired quantum numbers of the excited state.
Together with the reduction scheme for linearly dependent states
this guarantees the completeness of the $ph$ space when the
basis dimension goes to infinity. \\
In Fig.\ \ref{usualmit} the excitation spectrum for the
MIT-bag mean field obtained with different $ph$ interactions
is displayed. The excitation energies for various 
$ (J,T) $ channels are plotted as a function of the coupling
constant $G$. Note that the 
$ (\frac{3}{2}^+, \frac{3}{2}) $ channel was not considered
due to the specific problems discussed in part I. \\
As a global feature we observe that the states
are lowered in energy by the $ph$ interaction. In most cases
the decrease of the first excited state to zero energy occurs
within a narrow range of coupling constants. Eventually the
level meets its negative-energy counterpart (not shown) which has
been raised. The fact that the two levels meet at an energy
approximately equal to zero indicates that the spectrum
displays a symmetry under inversion of the energy axis. A closer
inspection of the whole spectrum including negative-energy
states shows that this symmetry is approximately preserved
until the first level is lowered to zero. A serious violation
of the symmetry has never been found in all the spectra we
have computed. \\  
The results for the color-current interaction are quite similar
to those for the $ \sigma -
\mbox{ \boldmath $ \pi $ \unboldmath } $ interaction when the
coupling constant is rescaled. As a difference the
color-current interaction is more repulsive in the channels
under investigation and the energy of some states is 
slightly increased. \\
Comparing these results to the experimental data we find
that for a standard MIT-bag configuration the level density
of $ph$ states is too high already on mean-field level
($ G=0 $). Whereas the absolute scale for the energies might not
be very well defined due to possible shifts of the ground-state energy
(which is lowered by correlations), the essential problem
lies in the small spacing between various states. \\
Even when the residual interaction is switched on the spectra
never resemble the experimental data for the baryon spectrum.
On the contrary, the level density gets further enhanced.
An eigenvalue which meets its negative-energy counterpart
at a certain value of $G$ develops a finite imaginary part
for larger values of the coupling constant. Only the real parts
are shown in the figures. \\  
Another global feature of the excitation spectra is the occurrence
of quasi-crossings. According to a theorem of Wigner and 
von Neumann \cite{LL77} levels belonging to the same representation
of a symmetry group cannot cross under the influence of a
perturbation. 
A closer inspection of the excitation spectra
obtained with the tensor-RPA method shows that some levels approach
up to a distance of a few $MeV$ but do not cross. \\
In order to obtain information on configuration mixing
in the spectrum we have studied the eigenvectors.
We only report some general features of the $ph$ amplitudes and
avoid displaying the vast amount of data. \\ 
When the levels are
well separated and ground-state correlations are negligible
there is always a dominant contribution of one $ph$ state
to the eigenvectors. Configuration mixing becomes sizeable
when the levels lie close together. \\
Ground-state correlations within an RPA approach are characterized
by nonvanishing $Y$-amplitudes. We find that 
for coupling constants up to values close to
the point where the first eigenvalue becomes degenerate
with the ground state the $Y$-amplitudes are negligible.
A characteristic feature of an MIT-bag and other stiff potentials
is the fact that the first excited state is lowered to zero
energy within a very narrow range of coupling constants.
Only within this range an eigenvector develops sizeable
$ Y $-amplitudes. The largest $ Y $-values are found for $ph$
excitations from the Dirac sea, reflecting the strong coupling
to negative-energy states due to a relatively small energy gap. 
Hence the level is driven down
by ground-state correlations originating from vacuum polarization. 
Ground-state correlations have been explicitly computed
in a simplified RPA scheme ignoring tensor coupling \cite{GHLM96}.
It has been found that such correlations are sizeable
and contributions from the Dirac sea are important. \\ 
The decrease of the first excited state in the manner of
a phase transition is a phenomenon well known in RPA theory. 
Model calculations \cite{SE73} indicate that this ``phase transition"
which is accompanied by an overestimate of ground-state
correlations is unphysical. With a refined approximation scheme
accounting for a more detailed description of the ground state 
(``self-consistent RPA") 
it can be shown that the ``phase transition" is replaced by
a smoother transition to low energies. The standard RPA
breaks down at the phase-transition point beyond which the assumptions
it is based on are no longer valid. However, we expect that 
before that point is reached important aspects of the excitation spectrum
as the existence of low lying collective states are 
correctly reproduced. \\   
As already mentioned, the negative-parity states
are expected to contain spurious admixtures. For this
reason they are not displayed here. In general the
mean-field excitation energy is lower for negative-parity 
states than for those of positive parity. When the
residual interaction is switched on the negative-parity
levels are driven to zero excitation energy
faster then the positive-parity levels, irrespective
of the $ph$ interaction which is used. This fact
stands in contradiction to the experimental data, where
the lowest excitation in, e. g., the 
$ (\frac{1}{2}^+, \frac{1}{2}) $ channel lies below that
of the $ (\frac{1}{2}^-, \frac{1}{2}) $ channel. \\ 
A similar problem with negative-parity states is encountered
in the nonrelativistic quark model \cite{GR95}.
Without perturbation the negative-parity excitations
lie well below those with positive parity. 
The problem can be cured by introducing
a nonrelativistic reduction of a meson-exchange interaction.
When such an interaction is treated perturbatively the order
of positive- and negative-parity states gets reversed.
In the relativistic case and allowing for
configuration mixing of $ph$ states,  
no such level
inversion is found, as can be deduced from our investigations
with the
$ \sigma - \mbox{ \boldmath $ \pi $ \unboldmath } $   
interaction. \\  
The question arises which feature of the relativistic 
RPA theory is responsible for
this deviation from the nonrelativistic case. 
In the nonrelativistic framework the level inversion
has a quite solid foundation. Within the constituent-quark
model its occurrence can be proven by purely group-theoretical
arguments while treating the meson-exchange interaction
in lowest-order perturbation theory \cite{GR95}.
When the three-body Faddeev equation is solved for the same
model the correct level ordering of positive- and negative-parity
states is still reproduced \cite{GPP96}. \\
In order to compare these calculations to our approach it is
helpful to disentangle the aspect of many-body approximation
schemes from the form of the interaction employed.
Since the spin-flavor structure of the interaction is
the keypoint of the level-ordering problem, we will
concentrate on this aspect exclusively. 
The $ \sigma - \mbox{\boldmath $ \pi $ \unboldmath } $ interaction
of Eq.\ \ref{NJL} does not reduce to the meson-exchange interaction
of Glozman et al.\ in the nonrelativistic limit.
An obvious difference is the $ \sigma $-exchange term which is
missing in the constituent-quark model. In addition to that
the so called tensor term arising from the nonrelativistic
reduction of a pion-exchange interaction \cite{EW88}
was neglected in Refs.\ \cite{GR95,GPP96}. \\
A suitable local interaction reducing in the non-relativistic limit 
to the structure used
by Glozman et al., however, is given by
\be
{\cal L}_{int} = \frac{G}{m^{*2}} \left( \nabla_i ( \Psi^{\dagger}
\Sigma_j \tau_k \Psi ) \right)^2 ,
\label{glozint} 
\ee
where the constituent-quark mass $ m^* $ is treated as a parameter
and replaces $m_0$ of Eq.\ \ref{colcu}.
$\mbox{\boldmath$\Sigma$\unboldmath}_i $ denotes the spin matrices
of Dirac theory and a summation over $ i,j,k $ is assumed.
The nonrelativistic limit of the spin-spin interaction  
Eq.\ \ref{glozint} is defined
via the lowest-order terms of the expansion in $ q/m^* $,
where a momentum-space representation of the interaction is 
used and $q$ denotes the momentum transfer. 
In lowest nonvanishing order of such an expansion
we obtain a two-body potential between quarks $i$ and $j$ 
with a spin-flavor structure
of the form $ \mbox{\boldmath$\sigma$\unboldmath}_i \cdot       
\mbox{\boldmath$\sigma$\unboldmath}_j \; \mbox{\boldmath$\tau$\unboldmath}_i
\cdot \mbox{\boldmath$\tau$\unboldmath}_j $ .
In spin-flavor space we thus recover a $ SU(2) $ version of
Glozman's meson-exchange interaction. \\ 
The expression of Eq.\ \ref{glozint} can be used as a $ph$ interaction
for the Dirac RPA problem in an equal manner as this was done
for different versions of the NJL model before. In order to
approach the nonrelativistic limit at least approximately without
completely blocking $ph$ admixtures from the Dirac sea
the constituent-quark mass was chosen $ m^* = 400 MeV $.    
When solving the tensor-RPA equation for the spin-spin interaction 
in the $ (\frac{1}{2}^+, \frac{1}{2}) $ and
$ (\frac{1}{2}^-, \frac{1}{2}) $ channel the lowest negative-parity
state is again found below the corresponding level with positive
parity. In these calculations the MIT-bag 
mean-field potential was used. \\
The same investigations were repeated in Tamm-Dancoff (TD) approximation.
The TD approximation amounts to solving a tensor-RPA
problem of half the dimension, keeping only $ M^{(1)} $ and
$ N^{(1)} $ in Eq.\ \ref{tensorrpa}. Physically, this means
that $ph$ correlations in the ground state are neglected.
This difference turns out to be crucial for the
ordering of positive- and negative-parity states.
In TD approximation the first excited state in the
$ (\frac{1}{2}^+, \frac{1}{2}) $ channel is found
below the corresponding state in the 
$ (\frac{1}{2}^-, \frac{1}{2}) $ channel when
the coupling constant $G$ is increased above a
critical value. \\ 
This observation leads us to suggest a simple
explanation for the wrong level ordering which is always obtained
in the RPA calculations. Even when the interaction
allows for a level inversion ground-state correlations
rapidly lower the first excited state and make it impossible
to further study the dynamical interplay of $ph$ configurations.
These ground-state correlations
come into the game well before the coupling constant
has reached the critical value corresponding to a level inversion
in TD approximation. As a result, the energy of the
first excited state in both parity channels is rapidly decreasing 
and no level inversion is observed. \\
The results so far discussed in this chapter should
be viewed as a case study which helps to shed light
on the range of applicability of the tensor-RPA
approach and and to explain typical features of the
excitation spectra. For this purpose a standard
MIT-bag mean-field configuration has been chosen.
However, the spectra do clearly not match with
experimental data and thus cannot be assigned
any physical significance. Considering the overabundance
of low-lying states we have obtained we could
only try to describe the baryon spectrum by introducing
a large number of unobserved states. \\
A more 
satisfactory approach is certainly to reduce the size
of the quark core. In combination with a comparatively
stiff mean-field potential this will effect a
shift of the excitation spectrum towards higher energies.
The size of the quark core which is needed to
fix the excitation energies to values in the
range of the experimental data is determined by
the $ T= \frac{3}{2} $ states. On mean-field level
the lowest $ T= \frac{3}{2} $ excitation appears at
the same energy as the corresponding $ T= \frac{1}{2} $
state. Experimentally, however, the energy of the
lowest $ T= \frac{3}{2} $ state tends to be
higher than the excitation energy in the 
$ T= \frac{1}{2} $ channel for the same $J^P$.
The $ \Delta(1232) $, which is not considered here,
forms an exception to this rule. Taking into account that
the $ph$ interaction in general effects a decrease of the
excitation energies, the mean-field energy gaps in the $ T= \frac{3}{2} $
channel have to exceed those of the lowest
excited state. This opens up the possibility to simultaneously
describe the $ T= \frac{1}{2} $ channel. In order to adjust
the mean-field spectrum correspondingly it is necessary to shrink
the size of the quark core to about $ 0.3 fm $. \\ 
The RPA spectra
obtained with the harmonic-oscillator potential
introduced in the previous chapter 
are shown in Fig.\ \ref{small}. For these investigations the color-current
interaction was used.
In order to compare these spectra with experimental data
the coupling constant has to be fixed. For this purpose
the $ (\frac{1}{2}^+, \frac{1}{2} )$ channel provides a suitable
guideline. In this channel the most rapid ``phase transition"
is observed, whereas for higher angular momenta the
phase-transition point is shifted to larger $ G $ (outside of the
$x$-range displayed in the figure).
Since beyond that point the RPA approximation breaks down
it is only meaningful to consider coupling constants
up to $ \approx 2.4 \cdot 10^{-6} MeV^{-2} $. On the
other hand, before the ``phase transition" the lowest
$ (\frac{1}{2}^+, \frac{1}{2} )$ excitation is too high in energy
whereas the mass splitting between the various states is
too low. We are thus naturally led to choose a coupling constant
of $ G = 2.35 \cdot 10^{-6} MeV^{-2} $ which allows us
to approximately adjust the $ (\frac{1}{2}^+, \frac{1}{2} )$
channel to the experimental data. At the same value of the
coupling constant the excitation energies in various other
channels are obtained as displayed in Fig.\ \ref{specfit}. \\ 
The horizontal bars indicate the energies obtained in the 
RPA calculation. The rectangles extend over the
range of uncertainty for the resonance maximum extracted from
$ \pi N $ scattering data \cite{PDG96}. The doubly hatched
areas represent $4*$ and $3*$ resonances, whereas the hatched
areas indicate the energy of $2*$ and $1*$ resonances.
Since some of the data for the latter are contradictory,
the corresponding range of energies has been extracted
from the most recent experiments in case of doubt. 
In order to obtain absolute numbers for the baryon masses
we have added the computed values for the excitation
energies to a nucleon mass of $ 940 MeV $. \\
Obviously the baryon spectrum is reproduced fairly well also
in the other channels. In the $ (\frac{5}{2}^+, \frac{1}{2} )$
and $ (\frac{7}{2}^+, \frac{3}{2} )$ channel we find low-lying
states which have not been observed in an experiment.
The residual interaction lifts the degeneracy of certain states
and effects a significant improvement over the mean-field
spectra. This statement, while
easily verified for the small-size quark core, is also true
in comparison with ground-state configurations with a larger rms. 
At the same value of $G$ the negative-parity excitations
are found at too low energies, as expected from 
spurious admixtures. \\
With the $ \sigma $-$ \mbox{ \boldmath $ \pi $ \unboldmath } $ 
interaction a comparable description of the baryon spectrum
could not be obtained. We found that this interaction tends to
lower the energy of all the levels in a very similar
fashion, thus preserving the structure of the mean-field
spectrum. We want to emphasize that for obtaining the spectra
we have not yet exhausted the full variational space of the
most general pointlike chiral interaction. In Ref.\ \cite{KLVW90}
such a variational procedure was undertaken for the meson spectra,
with the result that the color-current interaction Eq.\ 
(\ref{colcu}) provides a reasonable fit to the experimental data.
The baryon spectra presented here thus confirm the results
for the meson sector. In both of these calculations the
coupling constant $G$ enters as a parameter. In an NJL model
based on a self-consistent treatment of the ground state the
dynamics is governed by the dimensionless parameter $ G \Lambda^2 $.
Due to the lack of self-consistency in our approach the 
single-particle cut-off $ \Lambda $ is a priori not defined.
However, we can view the energy of the lowest-lying hole state included in
the RPA basis as a rough estimate of the 4-momentum 
scale under consideration. With this assumption we obtain
$ G \Lambda^2 \approx 15 $, compared to $ G \Lambda^2 = 10.51 $
of Ref.\ \cite{KLVW90}. Considering the approximate fashion of such
a comparison we conclude that the same coupling constant can be
used to describe the baryon and the meson sector. A natural extension
of our investigations with a simple prototype interaction would
be to consider the most general chiral interaction with independent
coupling strengths $ G_s $ and $ G_v $ for the scalar-pseudoscalar
and the vector-axialvector channels. This could improve the
description of the excitation spectrum. \\
We want to emphasize that $ q \bar{q} $ admixtures to
the wavefunctions are of considerable importance in the
RPA spectra. Although the valence energy obtained with the
harmonic-oscillator potential is quite high the $ph$ basis
contains more than twice as many excitations of the Dirac sea
as valence excitations. Clearly the lowest-lying and thus
most important states are the valence excitations. In some
cases, however, certain $ q \bar{q} $ amplitudes are of comparable
order of magnitude. Thus in spite of the large energy gap
the Dirac sea does not decouple. \\
The structure of the excitation spectrum is also sensitive
to the choice of the mean field. In comparison to the MIT-bag
results the transition of the levels to lower energies is
smoothed down, as seen in Fig.\ \ref{small}. This is due to
the relatively soft mean-field $ m(r) = c r^{\alpha} $ with
$ \alpha=2 $. This smoother transition extends the range of
coupling constants where RPA effects are important and results
in a simultaneous shift of excitation energies in several channels
when the $ph$ interaction is switched on. However, an even softer
parametrization $ \alpha=1 $ could provide a suitable
ground-state configuration only when the size of the quark core
is further reduced, as already pointed out in the previous chapter. \\
Although already a this stage a satisfactory description
of the baryon spectrum has been achieved, a few shortcomings
of our approach become apparent. In the
$ (\frac{1}{2}^+, \frac{1}{2} )$ channel the assumptions underlying
the RPA approach rapidly break down due to an overestimation
of ground-state correlations. In the $ T = \frac{3}{2} $ channels
the ``phase transition" occurs at slightly higher values of $G$. 
Initially the energy of the $ \Delta $
states with lowest angular momentum is slightly
increased. A shift of the $ \Delta $
states towards higher energies is expected from a color-current
interaction. The most prominent example of the effects of a
gluon-exchange force on the baryon spectrum has perhaps been given
in the context of the MIT-bag model \cite{DJ76}. At higher coupling
constants, however, this trend is overshadowed by 
rapidly growing ground-state correlations. \\
We expect that with a refined model for the ground state
including dynamical correlations the ``phase transition" is
smoothed down. Within such a scenario the $ T= \frac{3}{2} $
states could be lifted up more significantly, thus allowing
to employ a model with a slightly bigger quark core. 
Nevertheless our calculations already indicate that the nucleon
might contain a comparatively small quark core. The excitation
spectrum appears to be closely connected to the dynamics of
that quark core. It seems to be a reasonable approximation
to account for the meson cloud only in an implicit manner,
as inherent in a Dirac-RPA approach.
\section{Summary, Conclusion and Outlook}
The aim of this work was a phenomenological description
of the nucleon excitation spectrum in the framework of the
Dirac-RPA scheme developed in part I of this paper. \\
As an attempt towards a self-consistent treatment of the
mean-field configuration entering the RPA scheme we have
considered the Dirac-Hartree-Fock problem for a rotationally
and isospin invariant ground-state density.
For this purpose the Nambu-Jona-Lasinio model was used,
which is known to possess solitonic solutions in the $ B=1 $ sector. \\
In a calculation with the most general form of a self-energy
which is admissible within our treatment 
of symmetries the chiral condensate
of the NJL model could be reproduced with high accuracy.
The vacuum was found to be a translationally invariant state,
thus supporting the assumptions which usually enter the
Dyson-Schwinger equation for the vacuum self-energy. \\
The $ B=1$ system, however, did not emerge as a solitonic
object due to the unbound valence quarks. We conclude that the
pseudoscalar field, which does not contribute in our description
for reasons of symmetry, is of major importance for the
formation of the soliton found in other approaches \cite{ARW95,CGGP95}. 
The inclusion of a vector potential
does not cure the problem, as expected from the repulsive
character of the interaction. 
The results indicate that the existence of the soliton
does strongly depend on the grand-spin approach. \\
In order to allow for an application of the tensor-RPA method
and to construct a simple confinement mechanism
we have discussed a class of phenomenological models for the nucleon
ground state. In these descriptions confinement is enforced
by an asymptotically rising scalar potential. \\
We have solved the tensor RPA equation for different combinations
of mean-field potentials and $ph$ interactions with focus
on the positive-parity channels which are free of spurious
admixtures. It was found that by inclusion of a residual interaction
the excitation energies are lowered in most cases.
With increasing interaction strength strong ground-state
correlations develop. For a comparatively low valence energy
these correlations predominantly originate from the Dirac sea. \\
Using a standard MIT-bag configuration with a nucleon rms
of $ 0.71 fm $ the experimental data for the excitation 
energies cannot be reproduced.
The origin for this bad agreement lies in the high density of
excited states which is even increased when the $ph$ interaction
is switched on. \\
With regard to recent results obtained by Glozman and Riska 
\cite{GR95} we have also studied the problem of an inversion
of positive- and negative- parity states. It was found that in a
relativistic TD approximation the lowest positive- and negative-parity
states become inverted, whereas due to a strong
coupling to ground-state correlations no such inversion is seen
in the corresponding RPA calculation. \\
In view of reproducing the experimental data for positive-parity
baryons we have considered a quark core with a size
of $ \approx 0.3 fm $. At the same time the rigid MIT-bag
potential was replaced by a softer harmonic-oscillator potential.
By these means the baryon spectrum could
be reproduced fairly well, given that a local
color-current interaction is chosen, with a coupling strength
comparable to that of earlier work \cite{KLVW90}.
With a local $ \sigma - \mbox{\boldmath $\pi$ \unboldmath}$ interaction
no comparable description of the baryon spectrum could be achieved. \\
Despite of these successes some problems related to the
simplified description of the ground state remain. The RPA
``phase transition" due to an overestimation of ground-state
correlations effects a rapid breakdown of our method when
the coupling constant is increased above a critical value. 
By inclusion of dynamical correlations
in the ground-state wavefunction in an explicit fashion these
difficulties could be overcome. Since such a task is quite involved,
the corresponding refinements are left for future investigations. \\
Besides that two other issues clearly deserve a closer inspection.
On the one hand we have made no attempt to
improve the agreement with experimental data
by finetuning the independent
coupling constants $G_s$ and $G_v$ of a generalized chiral
interaction. On the other hand it is necessary to consider 
transition matrix elements in order to compute
the excitation probabilities of the various states in a $ \pi N $
scattering experiment. On such a basis one might be able to eliminate
certain states which are unobservable due to their reduced
transition strength. One will also be able to compute the partial
widths for, e. g., a decay into a $ N \pi $ final state.

\end{document}